\documentstyle[12pt]{article}
\begin{document}
\newcommand{\calm}{\cal M}
\newcommand{\calr}{\cal O}
\newcommand{\ma}{{\cal M}_a}
\newcommand{\mb}{{\cal M}_b}
\newcommand{\mbar}{\bar {\cal M}}
\newcommand{\mstar}{{\cal M}^{\star}}
\newcommand{\mup}{{\cal M}^{\uparrow}}
\newcommand{\mdown}{{\cal M}^{\downarrow}}
\newcommand{\tr}{{\rm {\bf {t\!r}}}}
\newcommand{\wkone}{\left| 1 \right\rangle _w}
\newcommand{\wktwo}{\left| 2 \right\rangle _w}
\newcommand{\bra}[1]{\left\langle #1 \right|}
\newcommand{\ket}[1]{\left| #1 \right\rangle}
\newcommand{\bkdot}[2]{\left\langle #1 \right.\left| #2 \right\rangle}
\newcommand{\bkdotw}{{ }_w\!\left\langle 1 \right.\left| 2 \right\rangle_w}
\newcommand{\bkdotwn}{\left|{ }_w\!\left\langle 1 \right.\left| 2
\right\rangle_w\right|}
\newcommand{\bkdotwns}{\left|{ }_w\!\left\langle 1 \right.\left| 2
\right\rangle_w\right|}
\newcommand{\ia}{{\cal O}_a}
\newcommand{\ib}{{\cal O}_b}
\newcommand{\iab}{{\cal O}_{ab}}
\newcommand{\iabbar}{{\cal O}_{\bar {ab}}}
\newcommand{\ih}{{\cal O}_H}
\typeout{--- Title page start ---}

\thispagestyle{empty}
\renewcommand{\thefootnote}{\fnsymbol{footnote}}

\begin{tabbing}
\hskip 11.5 cm \= {Imperial/TP/93-94/27}
\\
\> April, 1994 \\
\end{tabbing}
\vskip 1cm
\begin{center}
{\Large\bf Locating overlap information in quantum systems}
\vskip 1.2cm
{\large\bf Andreas Albrecht}\\
Blackett Laboratory, Imperial College\\
Prince Consort Road, London SW7 2BZ  U.K.\\
albrecht@ic.ac.uk
\end{center}
\vskip 1cm
\begin{center}
{\large\bf Abstract}
\end{center}

When discussing the black hole information problem the term ``information
flow'' is frequently used in a rather loose fashion.  In this article I
attempt to make this notion more concrete.  I consider a  Hilbert space
which is
constructed as a tensor product of two subspaces (representing for
example inside and outside the black hole).  I discuss how the system has
the capacity to contain information which is in {\em neither } of the
subspaces.  I attempt to quantify the amount of information located in  each of
the two subspaces, and elsewhere, and
analyze the extent to
which unitary evolution can correspond to ``information flow''.  I
define the
notion of ``overlap information'' which appears to be well suited to
the problem.

\vskip 1cm
\typeout{--- Main Text Start ---}

\renewcommand{\thefootnote}{\arabic{footnote}}
\setcounter{footnote}{0}

\section{Introduction}

The ``black hole information problem'' (BHIP) is one of the most
interesting topics in
theoretical physics.  At the root of this problem lie such fundamental
questions as ``is quantum gravity unitary?'', and ``can black holes decay
completely?'' (for an excellent review see
\cite{preskill93}).  The discussion often is phrased in terms of ``the flow of
information'' in and out of a black hole.

This is the first in a series of articles with the purpose of
answering the question
``what does information flow mean in quantum systems?''.  Is information some
locally conserved quantity which can
be followed as it ``flows'' from place to place?  If not, is it possible to
sensibly define information flow at all?

The motivation for this work is to understand what  constraints on
information flow  are implied
by unitary dynamics.   Because the preservation of all inner products is
the defining feature of unitary dynamics, I define ``overlap
information'', which allows one to reproduce the inner product of two
states.  It is this type of information which seems to be most
relevant to the problems considered here.

Aside from motivational aspects,  this first paper is not directly
concerned with
dynamics.   I
consider a Hilbert space which is constructed as a
tensor product of two subspaces.
I ask to what extent information can be said to be ``located in'' one
or the other of the spaces, and try to provide quantitative answers to this
question.   The constraints on this quantified
information implied by
unitary evolution are then examined.  A companion  paper will build on
this work and
discuss the dynamics of information flow in more detail.

The BHIP is the motivating force behind this work.  In section 2 I use
the BHIP to specify just what I mean by ``information''.
However, the
BHIP is quite complex and requires more than a definition of
information flow in order to be  resolved.    At this stage I cannot
claim that my efforts to quantify information flow offer any specific
new insights into the BHIP.  None the less, I do conclude that the
connections between unitarity and information flow are not simple ones.
It may prove productive to understand how the subtleties discussed here
apply to the BHIP.

This  paper is organized as follows: Section 2 provides the basic
framework for the discussion which follows.  Among other things, I
define what I mean
by information.  The definition I chose is
very narrowly motivated by the
BHIP and I call this information ``overlap information''.
There
purposely is  no
explicit reference to ``information theory'', or any other
of the various ways the term ``information'' shows up in the discussion of
physical problems.

Section 3 illustrates how  it is possible to hide information
from {\em both} of the subspaces.  I explicitly construct a
complete basis
for the whole tensor product space for which the information is completely
hidden from the subsystems.  That is, different basis vectors are
indistinguishable
from the point of view of either  subsystem.  I point out that many such
bases can be constructed.

Section  4 develops a way of quantifying the information in $a$ and
the information in $b$, and discusses the implications of unitary
evolution on these quantities.  In Section  5 I quantify the overlap
information
which is located elsewhere.

Section 6 follows up on some technical issues, and Section 7 contains
my conclusions.   Appendix A provides proofs for the
inequalities which I quote in the paper, and Appendix B provides a
concise list of the main definitions and inequalities from this paper.

\section{Preliminaries}

\subsection{What is  information?}

At the center of the BHIP is the question:  ``Does a decaying black
hole evolve in an entirely unitary fashion?''.  The hallmark of
unitary evolution is that the inner product between any two vectors in
the Hilbert space is preserved.

Hawking \cite{hawking75,hawking76} has argued that the process of
black hole decay can evolve two
orthogonal initial states into overlapping final states, and thus that
the decay process is
non-unitary.
This flies in the face of convictions which many physicists hold dear,
and Hawking's work has generated much heated debate.  Those who wish
to argue the case for a unitary decay process must show how
information about the initial state can be encoded in the decay
products (the ``Hawking radiation'') so that  orthogonal initial
states remain so over time.  If such a process can be identified, it
would be natural to say that information flows from inside to outside
the black hole.

In this work I take the point of view that the whole purpose of information
is to determine the inner product between two given states in the full
Hilbert space.   The question then is: Given any two states, where
does one have to look to acquire enough information to correctly
determine their inner product?  The answer, of course, will depend on
which two states one is working with.

Because one starts with two states, and uses the information to
discuss the relationship between these two state, I call the
information discussed here ``overlap information''.  The notion
appears to be quite different than other ways the term information is
used.
I stress that I wish to remain as focused as possible on issues
motivated by the BHIP.
Although I will sometimes drop the word ``overlap'', the term
``information'' will always refer to overlap information in what follows.

\subsection{The framework}

I will consider a ``world'' Hilbert space (denoted by $w$) which is
constructed as a tensor product of subspaces $a$ and $b$.  The sizes
of $a$ and $b$ are $N_a$ and $N_b$ respectively, and I will take $N_a
\leq N_b$.   I also take all spaces to be finite.

Given two
bases
\begin{equation}
\left\{ |i\rangle_a \right. \left| i = 1 ... N_a \right\}
\label{defbasisa}
\end{equation}
and
\begin{equation}
\left\{ |j\rangle_b \right. \left| j = 1 ... N_b \right\}
\label{defbasisb}
\end{equation}
which span subspaces $a$ and $b$ respectively, a tensor product  basis
which spans the
$w$ space is given by
\begin{equation}
\left\{\;\;  |i\rangle_a |j\rangle_b \;\; \right. \left| i = 1 ...
N_a, \; j = 1
... N_b \right\}.
\label{defbasisw}
\end{equation}
Any state $|\psi\rangle _w$ in $w$ can be written as
\begin{equation}
|\psi\rangle_w = \sum_{i,j} \alpha_{ij} |i\rangle_a |j\rangle_b.
\label{psiexp}
\end{equation}
where $i$ and $j$ run over their complete  ranges.

The results of any measurements of subsystem $a$ only can be
predicted using the density matrix given by
\begin{equation}
\rho_a \equiv \tr_b |\psi \rangle _w { }_w\!\langle \psi | \equiv
\sum_k \alpha^{\ast}_{ik}\alpha_{jk} |j\rangle_a { }_a\!\langle i |.
\label{rhoadef}
\end{equation}
The subsystem $a$ is only in a pure state if $\rho_a$ has but a single
non-zero eigenvalue.
Similarly, measurements on $b$ only can be predicted using $\rho_b$
produced by tracing over the $a$ subspace.  As long as $w$ is in a
pure state the non-zero eigenvalues of
$\rho_a$ and $\rho_b$ are always identical, and thus $a$ is in a pure
state if and only if $b$ is in a pure state.

Given two states $\wkone$ and $\wktwo$, one can construct the
corresponding $\rho_a^{(1)}$ and $\rho_a^{(2)}$.  If the value of
$\bkdotw$ can be correctly calculated using only $\rho_a^{(1)}$ and
$\rho_a^{(2)}$ I will say that the information resides in subsystem $a$.
Likewise one can construct $\rho_b^{(1)}$ and $\rho_b^{(2)}$, and if
the value of $\bkdotw$ can be calculated from these I will say that
the information lies in subsystem $b$.

I am not concerned in this paper with the question of what
measurements one actually has to do to extract the necessary
information from $a$ or $b$.  In many situation these may be
impossible for human beings to perform.  Despite such limitations, a
discussion of the information content of the $\rho_a$'s and $\rho_b$'s
seems a natural starting point for this work.  One must bear in mind
that in the end a discussion of the  BHIP will deal with
whether the fundamental Hamiltonian (not a human!) is able to move this
information about.

\subsection{The Schmidt expansion}
\label{tse}

It is often illuminating to expand a state $|\psi\rangle_w$ in the
``Schmidt'' basis \cite{s07,s35,z73,a92q}.  This is the tensor
product basis formed by using
the eigenstates of $\rho_a$ and $\rho_b$ to span their respective
subspaces.  Denoting the Schmidt basis by $S$:
\begin{equation}
|\psi \rangle_w = \sum_{j=1}^{N_a} \sqrt{p_j}
e^{i\phi_j}|j\rangle^S_a |j\rangle^S_b
\label{psischmidt}
\end{equation}
where $p_i$ are the non-zero eigenvalues which are all shared by the
two density matrices.   Note that the number of terms in the sum is
just the size of the smaller space ($N_a$) as opposed to $N_a\times
N_b$ in Eq \ref{psiexp}.  Each eigenstate of $\rho_a$ is
correlated with a unique eigenstate of $\rho_b$ with which it shares
the eigenvalue.
For a discussion of the Schmidt result see Appendix A in \cite{a92q}.

Normally the phase $\phi_j$ in Eq \ref{psischmidt} is absorbed into
the definition of the Schmidt basis states.  It is written explicitly
here because everything in Eq \ref{psischmidt} can almost always be
determined from
$\rho_a$ and $\rho_b$ except for this phase.  The eigenstates can be
determined up to a phase, and the correlations (that is which
eigenstates of $\rho_a$ and $\rho_b$
are paired together in the Schmidt  expansion) can be determined
because the correlated eigenstates share the same eigenvalue.

 The exception occurs
when two or more $p_i$'s are equal.  In this case the ``eigenvalue
matching'' just described  does not work in the subspace corresponding
to the degeneracy. One is thus unable to identify the correlations
within this subspace simply by knowing $\rho_a$ and $\rho_b$.
In the next section I shall take advantage of this
uncertainty to hide information from the subsystems.

\section{The hidden basis}

\subsection{Defining a hidden basis}

I will now construct a  special ``hidden'' basis which will serve to
illustrate some interesting places where information can be located.
The hidden basis is comprised of a {\em complete} set of orthonormal
states. For {\em each} element of the hidden basis $|i\rangle^H_w$ the
corresponding $\rho_a$ 's are all given by
\begin{equation}
\rho_a = {1 \over N_a} {\bf  I}_a
\label{diagrhoa}
\end{equation}
where ${\bf I}_a$ is the unit matrix in $a$.

Likewise, for each hidden basis state
\begin{equation}
\rho_b = {1 \over N_b} {\bf  I}_b.
\label{diagrhob}
\end{equation}
Complete hidden bases can only be constructed when $N_a = N_b$.
Since all the $\rho_a$'s and the $\rho_b$'s
are identical for all hidden basis states, one can not distinguish
among these states by examining the subsystems.

\subsection{The  $N_a = N_b =2$ case}
\label{tnaenbetwoc}

First I will take $N_a = N_b = 2$.  Using any bases denoted by
$\{ |1\rangle_a, |2\rangle_a\}$
and
$\{ |1\rangle_b, |2\rangle_b\}$
to span their respective subspaces the tensor product basis for $w$ is
\begin{equation}
\left\{
|1\rangle_a |1\rangle_b,\;\;
|1\rangle_a |2\rangle_b,\;\;
|2\rangle_a |1\rangle_b,\;\;
|2\rangle_a |2\rangle_b
\right\}.
\label{tpbasis}
\end{equation}

Now consider the first hidden basis state
\begin{equation}
|1\rangle^H_w \equiv {|1\rangle_a|1\rangle_b + |2\rangle_a|2\rangle_b
\over \sqrt{2} }.
\label{h1def}
\end{equation}
This state is written in Schmidt form, so one can immediately see that
$\rho_a = {1\over 2}{\bf  I}_a$ and
$\rho_b = {1\over 2}{\bf  I}_b$ as promised.

Now consider
\begin{equation}
|2\rangle^H_w \equiv {|1\rangle_a|1\rangle_b - |2\rangle_a|2\rangle_b
\over \sqrt{2} }.
\label{h2def}
\end{equation}
This state is also in Schmidt form and the subsystem density matrices
are identical to those corresponding to $|1\rangle^H_w$.  Even though ${
}^H_w\!\langle 1| 2\rangle^H_w = 0 $, one can not possibly deduce this
by inspecting $\rho_a$ and $\rho_b$.  This information is simply not
located in $a$ or $b$.  One can say that the information is located in
the relative {\em sign} of the two terms.
The remaining members of the hidden basis are
\begin{equation}
|3\rangle^H_w \equiv {|1\rangle_a|2\rangle_b + |2\rangle_a|1\rangle_b
\over \sqrt{2} }
\label{h3def}
\end{equation}
and
\begin{equation}
|4\rangle^H_w \equiv {|1\rangle_a|2\rangle_b - |2\rangle_a|1\rangle_b
\over \sqrt{2} }.
\label{h4def}
\end{equation}
These are constructed by interchanging  the correlations present in
$|1 \rangle^H_w$ and $| 2\rangle^H_w$ between the two subsystems.
The correlations can be changed without changing the $\rho$'s only
because the $p_i$'s are degenerate.

If one wants to reconstruct ${ }^H_w\!\langle 3| 1\rangle^H_w = 0 $
one could say that the information was ``in the correlations''
(whereas for ${ }^H_w\!\langle 4| 3\rangle^H_w = 0 $ the information
is again in the relative sign).

\subsection{Generalizing}
This construction can easily be generalized to arbitrary $N_a =  N_b
\equiv N$.
In this case ``information in the relative sign'' generalizes to
``information in the relative phases''.  One would have the first
$N_a$ hidden basis states being:
\begin{equation}
|n\rangle^H_w \equiv \sum_{k=1}^{N_a} e^{ikn2\pi/N}|k\rangle_a|k\rangle_b.
\label{hbphase}
\end{equation}
The remaining $N^2 - N$ hidden basis states can be constructed from
these by making all possible exchanges of correlations.

In the case where $N_b/N_a$ is an integer greater than unity one can
make a similar construction.  Since $\rho_b$ can have at most $N_a$
non-zero eigenvalues, the information can not be completely hidden
from the $b$ subsystem for all possible pairs of hidden basis
elements.  Some of the information will be contained in which
eigenvalues of $\rho_b$ are zero.

Note that many hidden bases can be constructed.  For every possible
choice of bases in the $a$ and $b$ subspaces one can construct a
different hidden basis in the manner illustrated above.  Still,  the
hidden basis states correspond to the special case where $\rho_a$ and
$\rho_b$  have
completely degenerate eigenvalues.  (It is interesting to note that
when $1 << N_a << N_b $ randomly chosen $|\psi\rangle_w$'s correspond
to $\rho_a \approx {\bf I}_a /N_a$ \cite{lubkin78,lloyd&pagles88,page93}.)

\section{Quantifying the information in $a$ and in $b$}

\subsection{The Schmidt perspective}

Consider now two arbitrary states $\ket{1}_w$ and $\ket{2}_w$. They
each can be Schmidt expanded giving
\begin{eqnarray}
\ket{1}_w & = & \sum_j \sqrt{p^{(1)}_j}  \exp({i\phi^{(1)}})\ket{j}^{(1)}_a
\ket{j}^{(1)}_b \label{sexpa} \\
\ket{2}_w & = & \sum_k \sqrt{p^{(2)}_k}  \exp({i\phi^{(2)}}) \ket{k}^{(2)}_a
\ket{k}^{(2)}_b .
\label{sexpb}
\end{eqnarray}
I have dropped the explicit $S$ superscript for Schmidt states.
Note that in general $\wkone$ and $\wktwo$ will generate {\em different}
Schmidt bases.  As discussed in section \ref{tse}, as long as the
$p_i$'s are not degenerate one can
construct Eqs \ref{sexpa} and \ref{sexpb} {\em except} for the $\phi$'s
just  from knowing $\rho_a$ and $\rho_b$.    For our purposes, the
utility of the Schmidt form lies in the ease with which it allows one
to identify what information about a state can be
accessed from the $\rho$'s.

\subsection{Measuring the overlap of two $\rho$'s}
It will be useful to define the quantity
\begin{equation}
\ma \equiv \tr_a \left( \sqrt{\rho_a^{(1)} } \sqrt{\rho_a^{(2)}}
\right) \equiv \sum_{jk} \sqrt{p_j^{(1)}p_k^{(2)}} \left| {
}_a^{(1)}\!\left\langle j|k
\right\rangle_a^{(2)} \right|^2.
\label{madef}
\end{equation}
This quantity only depends on $\rho_a^{(1)}$ and $\rho_a^{(2)}$. I
show in Appendix  A that  $\ma$ is
bounded by
\begin{equation}
0 \leq \ma \leq 1.
\label{mabounds}
\end{equation}
The quantity $\ma$ has the following properties:
\begin{equation}
\ma = 1 \;\;\;{\rm if} \;\;\; \rho_a^{(1)}= \rho_a^{(2)}
\label{ma1}
\end{equation}
\begin{equation}
\ma = 0 \;\;\;{\rm if} \;\;\; \rho_a^{(1)}  \perp  \rho_a^{(2)}.
\label{ma2}
\end{equation}
By $\rho_a^{(1)}  \perp  \rho_a^{(2)}$ I mean
that $\rho_a^{(1)}$ and $\rho_a^{(2)}$ assign
non-zero probabilities only to orthogonal subspaces.
 The  properties stated in Eqs \ref{ma1}  and \ref{ma2} are evident
from inspection of Eq \ref{madef}.

Equations \ref{ma1} and \ref{ma2} suggest that the value of $\ma$ gives a good
measure of
the ``overlap'' of $\rho_a^{(1)}$ with $\rho_a^{(2)}$.  This is how I
will use $\ma$.  (Note that using other powers of $\rho_a$ in Eq
\ref{madef} would not work. Check Eq \ref{ma1}  on $\rho_a^{(1)} =
\rho_a^{(2)} = {\bf I}_a/N_a$ to see this.)

Similarly one can describe the overlap between $\rho_b^{(1)}$ and
$\rho_b^{(2)}$  using
\begin{equation}
\mb \equiv \tr_b \left( \sqrt{\rho_b^{(1)} } \sqrt{\rho_b^{(2)}}
\right) \equiv \sum_{jk} \sqrt{p_j^{(1)}p_k^{(2)}} \left| {
}_b^{(1)}\!\left\langle j|k
\right\rangle_b^{(2)} \right|^2.
\label{mbdef}
\end{equation}

\subsection{The information in $a$}
\label{tiia}

In Appendix A I show that $\ma$ actually obeys
\begin{equation}
\bkdotwns \leq \sqrt{\ma}  \leq 1
\label{mabounds2}
\end{equation}
(which is more restrictive than Eq \ref{mabounds}).
This means that if $\ma = 0$ one knows for sure that $\bkdotw =0$.
When $\ma \neq 0$ Eq \ref{mabounds2} simply constrains $\bkdotwn$ to a
range of possible values.  In general, one is not able to constrain
$\bkdotwn$ any more tightly than
\begin{equation}
0 \leq \bkdotwns \leq \sqrt{\ma}
\label{bkdotwboundsa}
\end{equation}
if one only knows $\rho_a$.

As an illustration, suppose $\rho_a^{(1)} = \rho_a^{(2)} =
{\bf I}/N_a$.  Two possibilities are $|1\rangle_w = |1\rangle_w^H$ and
$|2\rangle_w = |2\rangle_w^H$, or  $|2\rangle_w = |1\rangle_w =
|1\rangle_w^H$ (where I am using the hidden basis vectors defined in
Section \ref{tnaenbetwoc}).  These give values of $\bkdotwn$ at opposite ends
of
the allowed range.

I offer here a {\em conjecture} (without proof)
that for {\em any} $\rho_a^{(1)}$ and $\rho_a^{(2)}$ one can choose
corresponding $\wkone$'s and $\wktwo$'s so that $\bkdotw$ lies anywhere in
the range given by Eq \ref{bkdotwboundsa}.

Recalling that for present purposes ``information'' corresponds to our
ability to determine $\bkdotw$, I define $\ia$, the ``overlap information
in $a$'' as
\begin{equation}
\ia \equiv {1 -\sqrt{\ma} }
\label{iadef}
\end{equation}

When $\ia = 1$ one is certain that $\bkdotw = 0$ and one can say
that all the information is in $a$.  When $\ia \neq 1$ one is less
certain of the value of $\bkdotw$, and there is incomplete information
in $a$.  When $\ia = 0$ there is no information in $a$ and $\bkdotw$
can take on {\em any} value ($0 \leq \bkdotwn \leq 1$).

Similarly, the information in $b$ is
\begin{equation}
\ib \equiv {1 -\mb }.
\label{ibdef}
\end{equation}

\subsection{Information flow between $a$ and $b$}

It is important to note that knowing $\ia$ alone tells us nothing
about $\ib$.  The information may be duplicated in both $a$ and $b$,
or there could be less in one than the other.  As is illustrated by
the hidden basis, it is quite possible that there is {\em no} information
in $a$ or $b$.  Table \ref{table1} gives some illustrations
\footnote{The density matrices  $\rho_a^{(1)}$ and $\rho_a^{(2)}$
contain more information than just $\ia$.  Under certain circumstances
it is possible to learn something about $\ib$ if  $\rho_a^{(1)}$ and
$\rho_a^{(2)}$ are known. For example, because $\rho_a$ and $\rho_b$
share the same eigenvalues, when $\rho_a \propto {\bf I}_a$ one can be
sure that $\rho_b \propto {\bf I}_b$ (as long as $N_a = N_b$).  Thus
(taking $N_a = N_b$)
if  $\rho_a^{(1)} = \rho_a^{(2)} \propto {\bf I}_b$ one can be certain
that $\ib = 0 $.  This point does not affect the discussion in this
section because with {\em other} forms for  $\rho_a^{(1)}$ and
$\rho_a^{(2)}$, $\ia$ and $\ib$ can be truly independent.
}.

Does unitarity alone constrain the evolution of $\ia$ and  $\ib$?
Not at all!  The only requirement unitarity imposes is that orthogonal
states must remain so over time. Given  two orthogonal initial
states, one can choose {\em any} pair of orthogonal final states and
construct a unitary time evolution which relates the two. Thus one can
construct unitary processes whereby $\ia$ and $\ib$ rise and fall
independently.  (Explicit examples of such constructions will appear in
\cite{albrecht&wandelt94}.)

Of course particular types of unitary evolution may produce a
relationship between $\ia$ and $\ib$, but the point I wish to make
here is that unitary evolution  alone does not imply ``information
flow'' between $a$ and $b$.  One can for example construct unitary
evolution where
$\ia$ evolves from unity to zero (so information ``flows out'' of
$a$), while $\ib$ holds steady at zero (so no information ``flows in''
to $b$).  For example, evolution from
columns  a) to column c)  in table \ref{table1} can do this (if $|{
}_a\!\langle X | Y\rangle_a| = 0$).

I have defined overlap information as that which allows you to evaluate the
inner product between two states.  Since unitary evolution preserves
inner products, surely there must be some kind of ``conservation of
overlap information'' associated with unitary evolution.  This is in
fact the
case.  The reason that the conservation of information does not force
information to flow between $a$ and $b$ is that there are {\em other}
places where the information might be located.  In what follows I will
try to quantify this information.

\section{Quantifying information which is neither in $a$ nor  $b$}

\subsection{Preliminaries}

As is amply illustrated by the hidden basis, it is possible for
overlap information to
be located in neither $a$ nor $b$.
Quantifying this information is not as straightforward as the
construction of $\ia$ and $\ib$.

My starting point is to use Eqs \ref{sexpa} and \ref{sexpb} to write
\begin{eqnarray}
\bkdotw & = &\sum_{jk} \sqrt{p^{(1)}_j p^{(2)}_k} \exp \left\{i(\phi_k^{(2)}
- \phi_j^{(1)})\right\}
\, { }^{(1)}_a \! \left\langle j| k\right\rangle^{(2)}_a
\, { }^{(1)}_b \! \left\langle j| k\right\rangle^{(2)}_b	\label{schdot} \\
& \equiv & \sum_{jk}M_{jk}e^{i\theta_{jk}}.
\label{mthetadef}
\end{eqnarray}
where
\begin{equation}
M_{jk} \geq 0.
\end{equation}
Equation \ref{schdot} is a sum over $N_a^2$ complex numbers.
Equation \ref{mthetadef} expresses  each of these complex numbers in
terms of its magnitude  $M_{jk} $ and complex phase $\theta_{jk}$.
Note that there in general are {\em fewer} $\phi$'s ($2N_a$) than
$\theta_{jk}$'s (which total $N_a^2$).  The values of the $\theta$'s
are in part determined
by the complex phases of the inner product of density matrix
eigenstates.   Thus, some information about the $\theta$'s can be
deduced from the $\rho$'s.  Barring eigenvalue degeneracy, the
values of the $M_{jk}$'s can be completely determined from the $\rho$'s.

One can visualize Eq \ref{schdot} as a ``chain'' in the complex plane.
Each term on the right side of Eq \ref{schdot} is represented by a
link in the chain with length $M_{jk}$.
The angle of orientation of each link is given by the complex phase
$\theta_{jk}$. The separation of the ends of the
chain equals
$\bkdotwn$.

In general there are $N_a^2$ links in the chain, but If  ${ }_w\!\bkdot{1}{2}_w
= 1$ then
\begin{equation}
M_{jk} = p_j\delta_{jk}
\label{m1}
\end{equation}
and
\begin{equation}
\theta_{jj} = 0 \;\;\; \forall j.
\label{zerotheta}
\end{equation}
Equation \ref{zerotheta} means that the chain is fully extended
in  the case of unit norm (all the links are parallel).  Equation \ref{m1}
shows
that in this  case the chain has at most $N_a$ links of
non-zero length.

\subsection{The length of the chain: $\mbar$}

Below I shall define some quantities for which no simple analytic
expression appears to exist.
For this reason it is helpful to start by
defining something which {\em does} have a simple form.  It will help
us get oriented, even though
in most cases it is not the most interesting quantity\footnote{
The quantity $\mbar$ is actually useful in one of the proofs in
Appendix A.
}.

I defined $\mbar$ by
\begin{equation}
\mbar \equiv \sum_{jk}M_{jk}.
\label{mbardef}
\end{equation}
This is just the {\em length} of the chain.
In  Appendix A I show that $\mbar$ obeys
\begin{equation}
\mbar \leq \sqrt{\ma \mb}.
\label{mbarleqmab}
\end{equation}
Also, the fact that the chain can not extend farther than its length
gives
\begin{equation}
\mbar \geq \bkdotwns .
\label{mbargeqbkdotwn}
\end{equation}
Taking Eqns \ref{mbarleqmab}, \ref{mbargeqbkdotwn}, and \ref{mabounds}
together gives
\begin{equation}
\left| \bkdotw \right| \leq \mbar  \leq \sqrt{\ma \mb}.
\label{mineq}
\end{equation}

Except in the case of eigenvalue degeneracy, The quantity $\mbar$ can
be calculated if one knows both $\rho_a$'s
and $\rho_b$'s, since then the only uncertainty lies in orientations
of the links, not their lengths.    In fact, since $\mbar$ is
completely independent of the link orientations, it actually fails to
make use of some information which {\em is } available if one one
knows both the  $\rho_a$'s
and $\rho_b$'s. (Remember, since in general there are more $\theta$'s
than $\phi$'s, the link orientations are not entirely independent of
the  $\rho_a$'s
and $\rho_b$'s.)  This is why $\mbar$ is not exactly the quantity I
would most like to evaluate.
When some eigenvalues are degenerate, knowing all the $\rho$'s will not be
enough to
determine $\mbar$.  (For an illustration of this point, compare
columns c) and d) in table \ref{table1}.)

\subsection{The hidden information: $\ih$}
\label{thirh}

Having defined $\mbar$ as a ``warm up'', I will now define quantities
which identify how much information is hidden from both $a$ and $b$.

First define $\mstar$ as the norm of the right hand side of Eq \ref{schdot}:
\begin{equation}
\mstar \equiv  \left| \sum_{jk} \sqrt{p^{(1)}_j p^{(2)}_k} \exp
\left\{i(\phi_k^{(2)}
- \phi_j^{(1)})\right\}
\, { }^{(1)}_a \! \left\langle j| k\right\rangle^{(2)}_a
\, { }^{(1)}_b \! \left\langle j| k\right\rangle^{(2)}_b \right|
\label{mstardef}
\end{equation}

Now define:
\begin{quote}
{\bf $\mup$}:  The maximum value $\mstar$ can achieve when the
$\phi$'s are allowed to vary arbitrarily.  When eigenvalues of a
$\rho$ are degenerate, the corresponding ambiguity in choice of
eigenstates (which appear in Eq \ref{mstardef}) must also be fully
explored as well, and the maximum value of $\mstar$ assigned to  $\mup$.
\end{quote}
and
\begin{quote}
{\bf $\mdown$}: The {\em minimum}  value of $\mstar$ achieved by
making the same variations as described in the definition of $\mup$.
\end{quote}
The hidden information $\ih$ is then
\begin{equation}
\ih \equiv \mup - \mdown.
\label{ihdef}
\end{equation}
Which represents the remaining uncertainty in $\bkdotwn$ once one
utilizes both the $\rho_a$'s {\em and } the $\rho_b$'s  to constrain
$\bkdotwn$ as tightly as possible.

\subsection{Putting it all together}

The quantity $\bkdotwn$ can lie anywhere between zero and unity.  To
the extent that one is able to constrain $\bkdotwn$ to lie in a more
narrow range $\Delta$ I will say that one has an amount of overlap
information given by $1 -
\Delta$.  This definition makes sense in the extreme limits of $\Delta
= 0$ and $\Delta = 1$, and implies a certain measure for intermediate
values (which I discuss in Section \ref{meas}).  My convention is that
when the overlap information equals unity one has complete
information.

Knowing $\ma$ (the overlap of the two $\rho_a$'s) allows one to
bound $\bkdotwn$ according to Eq \ref{mabounds2}.  Thus I arrived at
the ``overlap information in $a$'' defined by $\ia \equiv 1 -
\sqrt{\ma}$.

Knowing {\em both} $\ma$ and $\mb$ produces the constraint
\begin{equation}
0 \leq \left| \bkdotw \right| \leq  \sqrt{\ma \mb}
\label{mabbounds}
\end{equation}
so it is natural to define
\begin{equation}
\iab \equiv 1 - \sqrt{\ma \mb}.
\label{iabdef}
\end{equation}
But actually, if one knows  the $\rho_a$'s and the $\rho_b$'s one knows
a lot more than just $\ma$ and $\mb$.  One can employ the eigenvalue
matching described in Section \ref{tse} to determine the correlations.
Above I defined $\ih$ to be the information which remains hidden after
the $\rho_a$'s and $\rho_b$'s are used to maximal effect. Thus one can define
\begin{equation}
\iabbar \equiv 1 - \ih
\label{iabbardef}
\end{equation}
which is all the information one can possibly extract from the
$\rho_a$'s and the $\rho_b$'s.  When there is no eigenvalue degeneracy
one can say that $\iabbar - \iab$ is the information in the
correlations, and $\ih$ is the ``information in the phases''.  As we
have seen in the extreme case of the hidden
basis, when there is eigenvalue degeneracy some (or all) of the
information in the correlations can be hidden (and contribute to
$\ih$ instead of $\iabbar - iab$).

I show in Appendix A that (naturally enough)
\begin{equation}
\iabbar \geq \ia
\end{equation}
\begin{equation}
\iabbar \geq \ib.
\end{equation}

Table \ref{table1} gives some illustrations, where the ${\calm}$'s
and ${ \calr}$'s are given for a variety of $\wkone$'s and $\wktwo$'s.

\begin{table}[t]
\begin{tabular}{r||c|c|c|c|c}
 { } & a) & b) & c) & d) & e)  \\ \hline \hline
& & & & & \\
$\wkone$ &
$|X\rangle_a |U\rangle_b$ &
$|X\rangle_a |U\rangle_b$ &
$|1\rangle^H_w$ &
$|1\rangle^H_w$ &
$|1\rangle^H_w$
\\ \hline
& & & & & \\
$\wktwo$ &
$|Y\rangle_a |U\rangle_b$ &
$|Y\rangle_a |V\rangle_b$ &
$|2\rangle^H_w$ &
$|3\rangle^H_w$ &
$|1\rangle^H_w$
\\ \hline
& & & & & \\
$\bkdotwn$ &
$|{ }_a\!\langle X | Y \rangle_a|$ &
$|{ }_a\!\langle X | Y \rangle_a||{ }_a\!\langle U | V \rangle_a|$ &
$0$ &
$0$ &
$1$
\\ \hline \hline
& & & & & \\
$\sqrt{\ma}$ &
$|{ }_a\!\langle X | Y \rangle_a|$ &
$|{ }_a\!\langle X | Y \rangle_a|$ &
$1$ &
$1$ &
$1$
\\ \hline
& & & & & \\
$\sqrt{\mb}$ &
$1$ &
$|{ }_a\!\langle U | V \rangle_a|$ &
$1$ &
$1$ &
$1$
\\ \hline
& & & & & \\
$\sqrt{\ma \mb}$ &
$|{ }_a\!\langle X | Y \rangle_a|$ &
$|{ }_a\!\langle X | Y \rangle_a||{ }_a\!\langle U | V \rangle_a|$ &
$1$ &
$1$ &
$1$
\\ \hline
& & & & & \\
$\mbar$ &
$|{ }_a\!\langle X | Y \rangle_a|$ &
$|{ }_a\!\langle X | Y \rangle_a||{ }_a\!\langle U | V \rangle_a|$ &
$1$ &
$0$ &
$1$
\\ \hline
& & & & & \\
$\mup$ &
$|{ }_a\!\langle X | Y \rangle_a|$ &
$|{ }_a\!\langle X | Y \rangle_a||{ }_a\!\langle U | V \rangle_a|$ &
$1$ &
$1$ &
$1$
\\ \hline
& & & & & \\
$\mdown$ &
$|{ }_a\!\langle X | Y \rangle_a|$ &
$|{ }_a\!\langle X | Y \rangle_a||{ }_a\!\langle U | V \rangle_a|$ &
$0$ &
$0$ &
$0$
\\ \hline \hline
& & & & & \\
$\ia$ &
$1 - |{ }_a\!\langle X | Y \rangle_a|$ &
$1 - |{ }_a\!\langle X | Y \rangle_a|$ &
$0$ &
$0$ &
$0$
\\ \hline
& & & & & \\
$\ib$ &
$0$ &
$1 - |{ }_a\!\langle U | V \rangle_a|$ &
$0$ &
$0$ &
$0$
\\ \hline
& & & & & \\
$\iab$ &
$1 - |{ }_a\!\langle X | Y \rangle_a|$ &
$1 - |{ }_a\!\langle X | Y \rangle_a||{ }_a\!\langle U | V \rangle_a|$ &
$0$ &
$0$ &
$0$
\\ \hline
& & & & & \\
$\iabbar$ &
$1$ &
$1$ &
$0$ &
$0$ &
$0$
\\ \hline
& & & & & \\
$\ih$ &
$0$ &
$0$ &
$1$ &
$1$ &
$1$
\\ \hline
\end{tabular}
\caption{Some examples. Each column represents a particular choice of
$\wkone$ and $\wktwo$.  The values of the ${\calm}$'s and ${\calr}$'s
are given for each choice.  Note that here the hidden basis states
(with superscript $H$) refer to the $N_a = N_b = 2$ case discussed in
Section 3.2 .}
\label{table1}
\end{table}

\section{Further discussion}

\subsection{Relation to the Von Neumann entropy}

Given a state $|\psi\rangle_w$ and the corresponding $\rho_a$ and
$\rho_b$, the Von Neumann entropy $S$ (relative to the $a \otimes b$
partition) is given by
\begin{equation}
S \equiv - \tr_a \left(\rho_a \ln \rho_a\right) = - \tr_b \left(\rho_b
\ln \rho_b\right).
\label{entdef}
\end{equation}
If $|\psi\rangle_w$ has the pure  form
\begin{equation}
\left|\psi\right\rangle_w = \left| X \right\rangle_a \otimes \left| Y
\right\rangle_b
\label{pure}
\end{equation}
then $S = 0$.  When
\begin{equation}
\rho_a \propto {\bf I}_a/N_a
\label{maxent}
\end{equation}
 the entropy is
maximal (so $S = \ln(N_a)$)
\footnote{I am still taking (without loss of
generality) $N_a \leq N_b$.
}.

Given any two pure ($S=0$) states, {\em no} overlap information can be hidden
and $\ih = 0$.  The overlap information in $a$ and in $b$ ($\ia$ and
$\ib$) can each take on any value between zero and unity (see columns
a) and b) in table \ref{table1}).  In the
limit of two $S= 0$ states equality holds in Eq \ref{mineq} giving
\begin{equation}
\left| \bkdotw \right| = \mbar  = \sqrt{\ma \mb}.
\label{meq}
\end{equation}

In the other extreme, given any two states with maximal $S$, no
overlap information can be found in $a$, and if $N_b = N_a$ the same
holds for $b$ giving $\ia = \ib = 0$ and $\ih = 1$.  The hidden basis
states are an example of this.

Clearly the Von Neumann entropy has something to do with overlap
information.  Is it possible that the entire discussion
can be re-phrased
in terms of $S$, allowing one to avoid defining a new notion,
``overlap information'', as I have
done?  For example, Page \cite{page93} considers
a world partitioned in two, and uses $S$ to
discuss  the information in each of the two subsystems, as well as the
information in the correlations.

I believe it is clear that a discussion involving the Von Neumann
entropy of single states can not completely replace the notion of
overlap information.  For example if $N_b \geq 2N_a$ one can consider
pairs of orthogonal states with maximal entropy for which $\ib$ can take on
values anywhere between zero and unity (depending on the overlap of
the $\rho_b$'s).  Any discussion involving simply the Von Neumann
entropy could not distinguish among these cases.  Still, it may be
productive to try and understand more carefully the relationship
between Von Neumann entropy and overlap information.

\subsection{Measures}
\label{meas}

The ${\calm}$'s (and thus the ${\calr}$'s) which I have defined
place bounds on $\bkdotwn$.  I could
just as well placed bounds on $\bkdotwn ^{\alpha}$ for any real value
of $\alpha$.    For example, raising Eq \ref{bkdotwboundsa} to the
$\alpha$ power could result in
\begin{equation}
0 \leq \bkdotwns^{\alpha} \leq ({\ma})^{\alpha/2} \equiv \sqrt{\tilde
{\calm}} \equiv 1 - \tilde \ia .
\label{bkdotwboundsaalpha}
\end{equation}

Choosing a value of $\alpha$ other than unity would imply a different
measure of overlap information.  At this point I do not have a
concrete basis on which to distinguish among these different measures.
Things would be different if there were equalities relating sums of
${\calr}$'s (other than the rather trivial $\iabbar + \ih \equiv  (1 -
\ih) + \ih = 1$) which would not hold for the $\tilde {\calr}$'s I might
construct along the lines of Eq \ref{bkdotwboundsaalpha}.  I believe
the  fact
that such equalities do not exist is related to the lack of a clear
notion of information flow.  This is a result of the many places
overlap information can be located, and the possibility that
information can copied rather than being required to  flow {\em out}
of one place as it flows into another.

It is interesting to note that if one defines
\begin{equation}
m_a \equiv {1 \over 2}\ln (\ma)
\end{equation}
(and similarly $m_b$) then Eq \ref{mineq} becomes
\begin{equation}
 \ln\left| \bkdotw \right|   \leq m_a + m_b
\label{mineqln}
\end{equation}
which has an interesting additive form.  At this stage however, it is
not clear that there
are any real advantages to taking this path.

Another point is that if the conjecture in Section \ref{tiia} turns
out to be wrong, one probably would want to modify the choice of
measure.

\section{Conclusions}

I have defined the notion of ``overlap information'' which allows one
to try and relate the ``conservation of inner products'' corresponding
to  unitarity
with the notion of information flow.

I studied a space $w$ which is a
tensor product space of two subspaces: $w = a \otimes b$.  There are
four different places where the overlap information can be located:
In $a$, in $b$, ``in the phases'', and ``in the correlations''.  Given
two pure states in the $w$ space I have quantified these four types of
information in a way which seems useful, but is not unique.
Unitarity alone does not require information to execute a ``conserved
flow'', in the sense that information can flow {\em into} one location
without being required to flow out of another.

The existence of many complete ``hidden bases'' (for which {\em no}
information can be found in $a$ or $b$) means it is possible to start
with a complete set of initial states for which all the overlap
information is jointly held between $a$ and $b$, and unitarily evolve this
complete set into one in which the information is completely hidden
from $a$ and $b$.
Those who wish a unitary
resolution of the black hole information problem might want to look in
{\em all } possible  locations for the necessary
information.  In \cite{wilczek94} Wilczek takes a very interesting
step in this direction.

This article  has not discussed what type of Hamiltonians are required to
move information among these different locations.  Some work along
these lines is currently underway \cite{albrecht&wandelt94}.

\section{Acknowledgments}
I would like to thank J. Anglin, R. Laflamme, K. Stelle, B.
Wandelt, and W. Zurek  for helpful
discussions.   I am grateful for the hospitality of the Theoretical
Astrophysics group at Los Alamos National Laboratory, where this work
was completed.

\appendix

\section{Proofs of inequalities}

In this appendix I prove the various inequalities quoted in the paper.
They all
derive from the Schwarz inequality applied to suitably identified
real vector spaces with suitably defined inner products.  Everything I
do will be symmetrical under the exchange $a \leftrightarrow b$ so, in
particular, everything I prove for $\ma$ also applies to $\mb$.

\subsection{The Schwarz inequality}

If $V$ and $W$  are two vectors, {\em any} real quantity  $\langle V,W
\rangle$ is an  inner product if it obeys:
\begin{enumerate}
\item 	\begin{equation}
	\left\langle V,\left(\mu_1 W_1 + \mu_2 W_2\right)
	\right\rangle = \mu_1\left\langle V,W_1\right\rangle + \mu_2
	\left\langle V,W_2\right\rangle
	\label{bilinear}
	\end{equation}
	where $\mu_1$ and $\mu_2$ are real numbers.

\item  \begin{equation}
	\left\langle V,W \right\rangle = \left\langle W,V\right\rangle
	\label{symmetric}
	\end{equation}

\item	\begin{equation}
	\left\langle V,V\right\rangle \geq 0
	\label{posdef}
	\end{equation}
	with
	\begin{equation}
	\left\langle V,V\right\rangle =  0 \Rightarrow V = 0.
	\label{zerowhen}
	\end{equation}
\end{enumerate}

For any inner product, and any vectors $V$ and $W$ the Schwarz
inequality holds:
\begin{equation}
\left|\left\langle V,W \right\rangle\right|  \leq \left\langle V,V
\right\rangle
^{1/2} \left\langle W,W \right\rangle
^{1/2} .
\label{schwarz}
\end{equation}
Equality occurs when the $W$ and $V$ are linearly dependent.

Given any scalar product the norm $|| V ||$ of any vector can be
defined as $|| V || \equiv \left\langle V,V \right \rangle
^{1/2}$.  When working with vectors with unit norm,
the Schwarz inequality has  unity on the right hand side.

\subsection{ Bounds on $\ma$}

Equation \ref{madef} defines
\begin{equation}
\ma \equiv \tr_a \left( \sqrt{\rho_a^{(1)} } \sqrt{\rho_a^{(2)}}
\right) \equiv \sum_{jk} \sqrt{p_j^{(1)}p_k^{(2)}} \left| {
}_a^{(1)}\!\left\langle j|k
\right\rangle_a^{(2)} \right|^2.
\label{madefA}
\end{equation}

One can think of $\sqrt{\rho}$ as a  vector in a vector space with
inner product
\begin{equation}
\left\langle \sqrt{\rho}_V , \sqrt{\rho}_W \right\rangle \equiv
\tr\left( \sqrt{\rho}_V , \sqrt{\rho}_W  \right).
\end{equation}
(One can check that this meets all the requirements to be an inner
product)
Properly normalized density matrices (with unit trace) have unit
norm, and so
\begin{equation}
0 \leq \ma \leq 1.
\label{maboundsA}
\end{equation}
(Equation \ref{mabounds}) follows directly form the Schwarz
inequality and Eq \ref{posdef}.  I derive the  relationship between
$\ma$ and $\bkdotwn$ in section \ref{rmab}

\subsection{Bounds on $\mbar$}

The quantity $\mbar$ is defined by Eq \ref{mbardef} which, taken with
Eqs \ref{schdot} and \ref{mthetadef} gives
\begin{equation}
\mbar \equiv \sum_{j,k} \sqrt{p^{(1)}_{j} p^{(2)}_{k}}
\,\left| { }^{(1)}_a \! \left\langle {j}|
{k}\right\rangle^{(2)}_a \right|
\,\left| { }^{(1)}_b \! \left\langle {j}|
{k}\right\rangle^{(2)}_b\right|.
\label{mbardefA}
\end{equation}
It will be useful to think of Eq \ref{mbardefA} in terms of a single
sum by assigning to each unique ordered  pair $(j,k)$ a unique integer
$l(j,k)$.  The relation $l(j,k)$ can be inverted to give $j(l)$ and
$k(l)$.  Thus one can write
\begin{equation}
\mbar \equiv \sum_{l} \sqrt{p^{(1)}_{j(l)} p^{(2)}_{k(l)}}
\,\left| { }^{(1)}_a \! \left\langle {j(l)}|
{k(l)}\right\rangle^{(2)}_a \right|
\,\left| { }^{(1)}_b \! \left\langle {j(l)}|
{k(l)}\right\rangle^{(2)}_b\right|.
\label{mbardefA2}
\end{equation}

Now one can define
\begin{equation}
A_l \equiv  \sqrt[4]{p^{(1)}_{j(l)} p^{(2)}_{k(l)}}
\,\left| { }^{(1)}_a \! \left\langle {j(l)}|
{k(l)}\right\rangle^{(2)}_a \right|
\label{aldef}
\end{equation}
and
\begin{equation}
B_l \equiv  \sqrt[4]{p^{(1)}_{j(l)} p^{(2)}_{k(l)}}
\,\left| { }^{(1)}_b \! \left\langle {j(l)}|
{k(l)}\right\rangle^{(2)}_b \right|
\label{bldef}
\end{equation}
so $\mbar$ can be written as
\begin{equation}
\mbar = \sum_l A_l B_l \equiv \vec A \cdot \vec B.
\label{mbarabdef}
\end{equation}
Using this notation one finds that
\begin{equation}
\ma = \vec A \cdot \vec A
\end{equation}
and
\begin{equation}
\mb = \vec B \cdot \vec B.
\end{equation}
 One can identify $\vec A$ and $\vec B$  as vectors, and ``$\cdot$''
as a legitimate inner product.  The Schwarz  inequality then gives
\begin{equation}
\mbar \leq \sqrt{\ma  \mb}
\label{mbarschwarz}
\end{equation}

\subsection{Relating $\ma$ and $\bkdotwn$}
\label{rmab}

Equation \ref{mbarschwarz} implies a relation between $\ma$ and
$\bkdotwn$, even if $\mb$ is unknown.  One can simply use the fact
that $\mb$ is bounded above by unity (from Eq \ref{maboundsA}) to get
\begin{equation}
\bkdotwn \leq \sqrt{\ma}.
\label{mageqbkdotwn}
\end{equation}

Equality in Eq \ref{mbarschwarz} occurs when $\vec A \propto \vec B $.
Since the derivation of Eq \ref{mageqbkdotwn} is somewhat indirect,
one might wonder when (or even if) equality is achieved.  An
example where Eq \ref{mageqbkdotwn} becomes an equality appears in
column a) of Table \ref{table1}.

\subsection{Relating  $\iabbar$, $\ia$  and $\ib$}

The quantities $\iabbar$, $\mup$, and $\mdown$ are defined in Section
\ref{thirh} and Eq \ref{iabbardef}.
If there is no eigenvalue degeneracy, then
\begin{equation}
\mup \leq \mbar.
\label{mupleqmbar}
\end{equation}
This is because the maximization process which gives $\mup$ can not do
better than straightening out the chain, (giving $\mup = \mbar$) and
it could do worse (since the maximization process can not rotate all
links arbitrarily).  One thus has
\begin{equation}
\ih \equiv \mup - \mdown \leq \mup \leq \mbar \leq \sqrt{\ma \mb} \leq
\sqrt{\ma} .
\label{leq1}
\end{equation}
Multiplying Eq \ref{leq1} by $-1$, adding unity and using
\begin{equation}
\iabbar \equiv 1 - \ih
\end{equation}
and
\begin{equation}
\ia \equiv 1 - \sqrt{\ma}
\end{equation}
gives
\begin{equation}
\iabbar \geq \ia
\label{iabgeqia}
\end{equation}
Similarly one can show that
\begin{equation}
\iabbar \geq \ib.
\label{iabgeqib}
\end{equation}

If there {\em are} eigenvalue degeneracies it is possible to have
$\mup > \mbar$.  However, it is possible to construct an $\mbar '$ for
which $\mup \leq \mbar '$ and follow the above steps to arrive at Eqs
\ref{iabgeqia} and \ref{iabgeqib}.  This $\mbar '$ is constructed by
constructing new states $\wkone '$ and/or $\wktwo '$ by changing the
correlations present in the subspace corresponding to the degenerate
eigenvalues in order to maximize $\mbar$.  This maximum value of
$\mbar$ is $\mbar '$.  Since this maximization procedure does not
change $\ma$, $\mb$, or $\mup$ (which are the relevant quantities), the
above proof can be used with $\mbar '$.
Thus  Eqs
\ref{iabgeqia} and \ref{iabgeqib} apply even in the case of eigenvalue
degeneracy.

\section{Compilation of definitions and results}

Here is a concise compilation of definitions and results
which appear in this paper.

\subsection{The framework}
Two states, $\wkone$ and $\wktwo$, in a space $w$ partitioned
according to $w = a \otimes b$
may be Schmidt
expanded (Eqs \ref{sexpa} and \ref{sexpb}) to give
\begin{eqnarray}
\ket{1}_w & = & \sum_j \sqrt{p^{(1)}_j}  \exp({i\phi^{(1)}})\ket{j}^{(1)}_a
\ket{j}^{(1)}_b \label{sexpaB} \\
\ket{2}_w & = & \sum_k \sqrt{p^{(2)}_k}  \exp({i\phi^{(2)}}) \ket{k}^{(2)}_a
\ket{k}^{(2)}_b .
\label{sexpbB}
\end{eqnarray}

Each of $\wkone$ and $\wktwo$ generates its own $\rho_a$ and $\rho_b$,
resulting in $\rho_a^{(1)}$, $\rho_a^{(2)}$, $\rho_b^{(1)}$, and
$\rho_b^{(2)}$.

\subsection{Overlap and information in $a$}
The overlap of $\rho_a^{(1)}$ and   $\rho_a^{(2)}$ is given by Eq \ref{madef}:
\begin{equation}
\ma \equiv \tr_a \left( \sqrt{\rho_a^{(1)} } \sqrt{\rho_a^{(2)}}
\right) \equiv \sum_{jk} \sqrt{p_j^{(1)}p_k^{(2)}} \left| {
}_a^{(1)}\!\left\langle j|k
\right\rangle_a^{(2)} \right|^2
\label{madefB}
\end{equation}
which obeys (Eq \ref{mabounds2})
\begin{equation}
\bkdotwns \leq \sqrt{\ma}  \leq 1.
\label{mabounds2B}
\end{equation}
This leads to the definition of $\ia$  the ``overlap information in
$a$'':
\begin{equation}
\ia \equiv {1 -\sqrt{\ma} }
\label{iadefB}
\end{equation}
(Eq \ref{iadef}).
The equivalent quantities can be defined for $b$ (Eqs \ref{mbdef} and
\ref{ibdef}).

\subsection{The quantity $\mbar$}
The quantity $\mbar$ is defined by Eqs \ref{mbardef}, \ref{schdot}, and
\ref{mthetadef}:
\begin{equation}
\mbar \equiv \sum_{j,k} \sqrt{p^{(1)}_{j} p^{(2)}_{k}}
\,\left| { }^{(1)}_a \! \left\langle {j}|
{k}\right\rangle^{(2)}_a \right|
\,\left| { }^{(1)}_b \! \left\langle {j}|
{k}\right\rangle^{(2)}_b\right|
\label{mbardefB}
\end{equation}
which obeys
\begin{equation}
\left| \bkdotw \right| \leq \mbar  \leq \sqrt{\ma \mb}
\label{mineqB}
\end{equation}
(Eq \ref{mineq}).

\subsection{The hidden information}
The quantity $\mstar$ is given by (Eq \ref{mstardef})
\begin{equation}
\mstar \equiv  \left| \sum_{jk} \sqrt{p^{(1)}_j p^{(2)}_k} \exp
\left\{i(\phi_k^{(2)}
- \phi_j^{(1)})\right\}
\, { }^{(1)}_a \! \left\langle j| k\right\rangle^{(2)}_a
\, { }^{(1)}_b \! \left\langle j| k\right\rangle^{(2)}_b \right|
\label{mstardefB}
\end{equation}

The quantities $\mup$ and $\mdown$  are:
\begin{quote}
{\bf $\mup$}:  The maximum value $\mstar$ can achieve when the
$\phi$'s are allowed to vary arbitrarily.  When eigenvalues of a
$\rho$ are degenerate, the corresponding ambiguity in choice of
eigenstates (which appear in Eq \ref{mstardef}) must also be fully
explored as well, and the maximum value of $\mstar$ assigned to  $\mup$.
\end{quote}
and
\begin{quote}
{\bf $\mdown$}: The {\em minimum}  value of $\mstar$ achieved by
making the same variations as described in the definition of $\mup$.
\end{quote}
The hidden information $\ih$ (Eq \ref{ihdef}) is then
\begin{equation}
\ih \equiv \mup - \mdown
\label{ihdefB}
\end{equation}
which represents the remaining uncertainty in $\bkdotwn$ once one
utilizes both the $\rho_a$'s {\em and } the $\rho_b$'s  to constrain
$\bkdotwn$ as tightly as possible.

\subsection{The information in {\em both} $a$ and $b$}
The most overlap information one can extract from  the  $\rho_a$'s
{\em  and} $\rho_b$'s  is
\begin{equation}
\iabbar \equiv 1 - \ih
\label{iabbardefB}
\end{equation}
(Eq \ref{iadef}).
Naturally,
\begin{equation}
\iabbar \geq \ia, \;\; \ib.
\end{equation}
The overlap information in both  $\ma$ and $\mb$ is
\begin{equation}
\iab \equiv 1 - \sqrt{\ma \mb}.
\label{iabdefB}
\end{equation}
When there is no eigenvalue degeneracy
one can say that $\iabbar - \iab$ is the information in the
correlations, and $\ih$ is the ``information in the phases''.
When there is eigenvalue degeneracy some (or all) of the
information in the correlations can be hidden (and contributes to
$\ih$ instead of $\iabbar - \iab$).

\end{document}